\documentclass[fp, twocolumn]{jpsj3}
\usepackage{txfonts}
\usepackage{amsmath, amssymb}
\usepackage{bm}
\usepackage{cite}
\usepackage{graphicx, color}
\title{Open Quantum Dynamics Theory of Spin Relaxation: Application to $\mu$SR and Low-Field NMR Spectroscopies}

\author{Hideaki Takahashi\thanks{Email: hideaki@qchem.kuchem.kyoto-u.ac.jp} and Yoshitaka Tanimura\thanks{Email: tanimura.yoshitaka.5w@kyoto-u.jp}}
\inst{Department of Chemistry, Graduate School of Science, Kyoto University, Kyoto 606-8502, Japan} 

\abst{An open quantum system is a system coupled to an environment that can describe time-irreversible dynamics through which the system evolves toward the thermal equilibrium state. We present a quantum mechanically rigorous theory in order to help the analysis of spectra obtained from the advanced nuclear magnetic resonance (NMR) and muon spin rotation, relaxation or resonance ($\mu$SR) techniques. Our approach is based on the numerically ``Exact'' hierarchical equations of motion (HEOM) approach, which allows us to study the reduced system dynamics for non-perturbative and non-Markovian system--bath interactions at finite temperature even under strong time-dependent perturbations. We demonstrate the present theory to analyze $\mu$SR and low-field NMR spectra, as an extension of the Kubo--Toyabe theory focusing on the effects of temperature and anisotropy of a local magnetic field on spectra, to help further the development of these experimental means.}

\begin{document}
\maketitle
\thispagestyle{empty}

\section{Introduction}
For the analysis of NMR and ESR spectroscopies, the quantum master equation or the Redfield theory has been developed to describe the effects of the longitudinal and transversal relaxations characterized by the time constants $T_1$ and $T_2$.\cite{Bloch53, Redfield65} Then, the stochastic theory has been employed to describe the effects of the inhomogeneous dephasing characterized by the time decay constant $T_2^{\dag}$ in the fast modulation limit.\cite{Kubo69}  Owing to the advent of experimental techniques that include NMR and ESR, spin dynamics are now investigated under extreme physical conditions, such as quantum computing, where the quantum nature of an environment plays an essential role.\cite{Nishimori98, Dwave}  Thus, such existing theories are insufficient to investigate the complex motion of a spin system. This is also true for zero- to ultralow-field NMR measurement\cite{Guenget2013NMR, Ledbetter2011, Sjolander2017, Barskiy2019} and muon spin rotation, relaxation or resonance ($\mu$SR) spectroscopy, \cite{Yaouanc2011} because the excitation energy of a spin is almost zero in such measurements and quantum thermalization processes play an important role even at very low temperatures.

$\mu$SR spectroscopy is a magnetic resonance technique that utilizes a short-lived elementary particle, a muon (lifetime: 2.2 $\times 10^{-6}$ s). The muon is a charged spin particle whose magnetic moment is three times larger than that of a proton.  Because of its large magnetic moment and short lifetime, it can be implanted in matter to obtain extremely sensitive local magnetic and electronic probes. $\mu$SR spectroscopy measures the muon spin polarization recorded from the decay anisotropy of the emitted positrons, as a function of the arrival and decay times. While the experimental setup is completely different from that of NMR spectroscopy, the information obtained by $\mu$SR spectroscopy is analogous to that by a low-field NMR measurement.

In 1966, Kubo and Toyabe developed the spin relaxation theory for NMR in zero or weak external magnetic field comparable to the local field from a stochastic approach.\cite{kubo-toyabe1967}. Such a low-field measurement was then realized by $\mu$SR spectroscopy, and since then, the Kubo-Toyabe theory has been employed to analyze the long-time behavior of the $\mu$SR spectrum to probe a local environment of materials.\cite{hayano1979, uemura1980, kubo1981, kubo1987,aoyama1992, aoyama1993}
Various materials that include itinerant helimagnets, superconductors, proteins and DNA have been studied by $\mu$SR spectroscopy.
\cite{kadono1989, kadono1990, Matsuda1997, Sonier2001, Nagamine2000, Torikai2006, Takeshita2009, Hiraishi2014}
While several theories for $\mu$SR spectroscopy have been developed,\cite{celio-meier1983, celio1986, sonier2012, keren1994, crook1997} the Kubo-Toyabe theory is commonly used for investigations of this kind, because it is handy while describing the experimentally obtained $\mu$SR signal reasonably well.
This feature arises from the assumption that the three-dimensional local random field surrounding the muon is described by a stochastic noise, $\Omega_{\alpha}(t)$ for $\alpha = x, y, z $, which undergoes the Gaussian--Markovian process determined by the noise correlation function, $\left\langle \Omega_{\alpha}(t+t_0) \Omega_{\alpha}(t_0) \right\rangle ={\Delta^2} \mathrm{e}^{-\nu t}$, where $\Delta$ and $\nu$ are the amplitude and inverse correlation time of the noise, respectively. This allows us to employ the stochastic Liouville equation (SLE) to describe the spin dynamics of the muon. This equation can be solved analytically in a continued fractional form; the static limit of the spin relaxation function is now called the Kubo-Toyabe function.

Although the Kubo-Toyabe theory is convenient to use, there are many limitations in applying it to the analysis of experimental results. For example, this theory does not account for a temperature effect, because the stochastic theory is phenomenological and cannot describe the thermal equilibrium state at finite temperature. It is also applicable only to an isotropic environment without any external forces. Several improvements have been made in the framework of the stochastic theory,\cite{hayano1979, uemura1980, kubo1981, kubo1987, aoyama1992, aoyama1993, keren1994, crook1997} but applicability is still limited. This is because the stochastic theory relies on the Markovian assumption, whereas the local noise that we investigate arises from the non-Markovian vibrational motion of inter- and intra-atomic or molecular modes in a complex material.

To eliminate the above-mentioned limitations, here, we consider a system-bath model to treat the system dynamically and use the numerically ``exact'' hierarchical equations of motion (HEOM) approach to calculate spectra in a rigorous manner.\cite{tanimura1989,TanimuraPRA90,TanimuraPRA91,ishizaki2005, ishizaki2006,tanimura2006,Tanimura2014,Tanimura2015} The HEOM are the equations of motion that can describe the dynamics of a system for non-perturbative and non-Markovian system--bath interactions at any temperature. In the high temperature limit, the HEOM results for the Drude bath spectral distribution agree with those from the stochastic theory: The HEOM can be regarded as a generalization of the SLE. Most importantly, the HEOM have flexibility to take into account the effects of a realistic noise that can be obtained from experimental means or molecular dynamics simulations.

This paper is organized as follows. In Sect. \ref{theory}, we present a typical model system for the NMR and $\mu$SR spectroscopy analyses. The HEOM and their characteristic features are described. Numerical results and discussion are presented in Sect. \ref{Results}.  Section \ref{conclusion} is devoted to our conclusions.

\section{Theory}
\label{theory}
\begin{figure}[t]
  \centering
  \includegraphics[scale = 0.2]{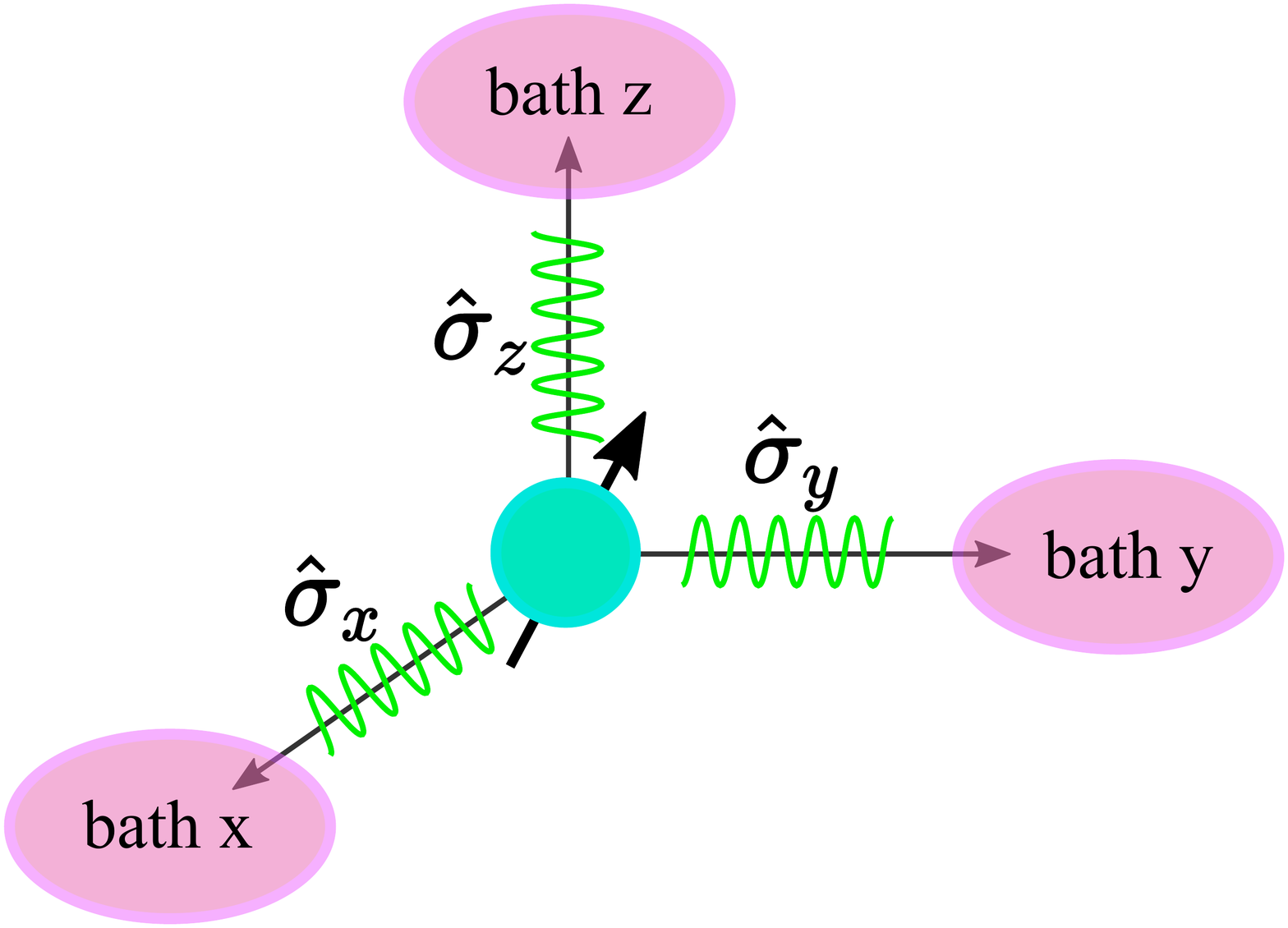}
  \caption{(Color online) Schematic depiction of three-dimensional spin-Boson model for $\mu$SR spectroscopy.}
  \label{f1}
\end{figure}

\subsection{Spin-Boson model in three-dimensional space}
We consider a spin system as a probe of a local magnetic environment for $\mu$SR and NMR spectroscopies described by
\begin{equation}
  \hat{H}_{\mathrm{S}} = - \frac{1}{2} \hbar \omega_{0} \hat{\sigma}_{\mathrm{z}} - \frac{1}{2} \hbar \hat{\bm \mu} \cdot  {\bm B}(t),
\end{equation}
where
$\hat{\bm \mu}\equiv \mu_0 \left( \sin\theta \cos\phi \hat{\sigma}_{x}, \sin\theta \sin\phi \hat{\sigma}_{y}, \cos\theta \hat{\sigma}_{z} \right)$ is the dipole operator with the amplitude $\mu_0$ expressed as a function of the solid angle, and  $\hat{\sigma}_{\mathrm{\alpha}}$ ($\alpha = x, y,$ and $z$) are Pauli matrices.  The frequency $ \omega_{0}$ is the Zeeman energy that arises from the static longitudinal external field, which is set to zero in the case of zero-field $\mu$SR and NMR spectroscopies. The function ${\bm B}(t) $ is the time-dependent external field introduced to describe various experimental schemes, which include spin echo, COrrelation SpectroscopY (COSY), and Nuclear Overhauser Effect SpectroscopY (NOESY) measurements utilizing $\pi$ and/or $\pi/2$ pulses.\cite{Guenget2013NMR} Using ${\bm B}(t) $, we can explicitly treat nonthermal vibrational motion that is, for example, evaluated from molecular dynamics simulations.

The spin system is independently coupled to three heat baths in the $x$, $y$, and $z$ directions to describe an environment in a three-dimensional space (see Fig. 1). We can regard these baths as arising from a local magnetic field owing to the surrounding atoms or molecules. The total Hamiltonian is then given by
 \begin{equation}
  \hat{H}_{\mathrm{tot}} = \hat{H}_{\mathrm{S}} + \sum_{\alpha = x, y, z} \left(\hat{H}_{\mathrm{I}}^{\alpha} + \hat{H}_{\mathrm{B}}^{\alpha} \right),
\end{equation}
where
\begin{equation}
  \hat{H}_{\mathrm{B}}^{\alpha} = \sum_{j} \hbar \omega_j^{\alpha} \left( \hat{b}_j^{\alpha \dagger} \hat{b}_j^{\alpha}  + \frac{1}{2} \right),
\end{equation}
and
\begin{equation}
  \hat{H}_{\mathrm{I}}^{\alpha} = \hbar \hat{V}_{\alpha} \sum_{j} g_j^{\alpha} \left(\hat{b}_j^{\alpha \dagger} + \hat{b}_j^{\alpha} \right),
\label{SBINT}
\end{equation}
and $\hat{H}_{\mathrm{B}}^{\alpha}$ and $\hat{H}_{\mathrm{I}}^{\alpha}$ are the Hamiltonian of the $\alpha$th bath and the Hamiltonian representing the interaction between the system and the $\alpha$th bath, respectively.
The system part of the system-bath interactions is defined as $\hat{V}_{\alpha} =  \hat{\sigma}_{\alpha}/2$, and $ \hat{b}_j^{\alpha}, \hat{b}_j^{\alpha \dagger}, \omega_j^{\alpha}$, and $g_j^{\alpha}$ are the annihilation operator, creation operator, frequency, and system--bath coupling constant for the $j$th mode of the $\alpha$th bath, respectively. For conventional NMR measurements, we consider the heat bath in the $z$ direction only, because the effects of the noise in the $x$ and $y$ directions can be ignored owing to the large $\omega_0$.

The $\alpha$th heat bath can be characterized by the spectral distribution function (SDF), defined by
\begin{equation}
  J_{\alpha}(\omega) = \sum_{j} (g_j^{\alpha})^{2}\; \delta(\omega - \omega_j^{\alpha}).
\end{equation}
By adjusting the form of the SDF, the properties of the local environment consisting of solid--state materials, solvates, and protein molecules can be modeled. The SDF is estimated from spectroscopic experiments\cite{IR_exp1,Palese96, Pullerits20} or simulations.\cite{Renger06, KramerAspu13, KramerAspu14, Coker2016, Ueno2020}
If we reduce the bath degrees of freedom to obtain the reduced density matrix $\hat{\rho}_\mathrm{S} (t)=\mathrm{tr}_\mathrm{B} \{ \hat{\rho}_{\mathrm{S+B}} (t)\}$, the baths produce the noise on the system defined as $\hat{\Omega}_{\alpha}(t) \equiv \sum_{j} g_{j}^{\alpha} \hat{x}_{j}^{\alpha}(t)$. Through these noise terms, the bath thermalizes the system through fluctuation and dissipation. For a harmonic bath, the effects of thermal fluctuation are expressed as the symmetrized correlation function defined by\cite{tanimura1989,tanimura2006}
\begin{align}
  \frac{1}{2}\left\langle \left\{ \hat{\Omega}_{\alpha} (t), \hat{\Omega}_{\alpha} (0) \right\} \right\rangle_{\mathrm{B}}
  = \int \mathrm{d} \omega J_{\alpha}(\omega) \coth\left( \frac{\beta\hbar\omega}{2} \right) \cos(\omega t) ,
  \label{fluctuation}
\end{align}
whereas that of the relaxation function is expressed as the anti-symmetrized correlation defined by
\begin{equation}
  i\left\langle \left[\hat \Omega_{\alpha} (t), \hat \Omega_{\alpha}(0) \right] \right\rangle_{\mathrm{B}} = i\int d\omega J_{\alpha}(\omega) \sin( \omega t),
\label{dissipation}
\end{equation}
where $\langle \cdots \rangle_\mathrm{B}$ represents the thermal average of the bath degrees of freedom. The symmetrized and anti-symmetrized correlation functions relate through the quantum fluctuation-dissipation theorem. The relationship between the present dynamical theory and the stochastic theory can be illustrated using the classical Langevin equation that can be derived from the system-bath model.\cite{tanimura2006} In the Langevin approach, the function $\hat{\Omega}_{\alpha}(t)$ corresponds to the Langevin random force whose correlation function is defined by Eq. \eqref{fluctuation}. The damping kernel of the Langevin equation is then expressed as Eq. \eqref{dissipation}, which relates with Eq. \eqref{fluctuation} through the classical fluctuation-dissipation theorem. The stochastic theory thus corresponds to the Langevin formalism without the damping term, because the theory ignores the effects of dissipation. Such a situation is only true when the bath temperature is extremely high and the damping kernel becomes smaller than the fluctuation term. Because the HEOM formalism treats both fluctuation and dissipation, it can describe the irreversible dynamics of the system accurately, whereas the stochastic theory describes the dephasing motion only.

In principle, the HEOM can be constructed for any profile of SDF for the noise correlation functions expressed in terms of damped oscillators as $\exp[-\zeta^{a} \pm i \omega^{a}]$, where $\zeta^{a}$ and $\omega^{a}$ characterize the relaxation and oscillation of the noise correlation for the $\alpha$th bath, respectively: The HEOM have been derived for the Drude, \cite{tanimura1989,TanimuraPRA90,TanimuraPRA91,ishizaki2005, ishizaki2006,tanimura2006,Tanimura2014,Tanimura2015} Brownian,\cite{TanimuraMukamelJPSJ94,TanakaJPSJ09, TanakaJCP10} Lorentz,\cite{Nori12} Ohmic,\cite{Ikeda2019Ohmic} Drude-Lorentz,\cite{KramerFMO2DLorentz} and their combinations\cite{TanimruaJCP12, KramerJPC2013BO2DEcho}. Alternatively, we can derive the HEOM for an arbitrary SDF using the Chebyshev-quadrature spectral decomposition to study the sub-Ohmic SDF at zero temperature.\cite{WueHEOM2017}  The HEOM for Brownian spectral distribution, which is important to take into account local modes of an environment, are presented in Appendix.

Here, to demonstrate the relationship between the HEOM approach and the stochastic approach, we consider the Drude SDF defined by\cite{tanimura1989,TanimuraPRA90,TanimuraPRA91,ishizaki2005, ishizaki2006,tanimura2006,Tanimura2014,Tanimura2015}
\begin{equation}
  J_{\alpha}(\omega) = \frac{\eta_{\alpha}}{\pi} \frac{\gamma_{\alpha}^{2} \omega}{\gamma_{\alpha}^{2}+\omega^2 },
\end{equation}
where  $\eta_{\alpha}$ represents the coupling strength between the system and the $\alpha$th bath. If necessary, we define $\eta_{\alpha}$ as a function of solid angle to represent the rotationally invariant environment.\cite{Iwamoto2018,Iwamoto2019} The correlation functions are then analytically evaluated as \cite{tanimura1989,TanimuraPRA90}
\begin{equation}
  \frac{1}{2}\left\langle \left\{ \hat \Omega_{\alpha} (t), \hat \Omega_{\alpha} (0) \right\} \right\rangle_{\mathrm{B}}  =  \sum_{k=0}^{\infty} c_k^{\alpha} \mathrm{e}^{-\nu_k ^{\alpha} |t|},
  \label{fluctuation1}
\end{equation}
and
\begin{equation}
  i \left\langle \left[ \hat{\Omega}_{\alpha}(t), \hat{\Omega}_{\alpha}(0) \right] \right\rangle_{B} = 2{\bar c}_0^{\alpha} \mathrm{e}^{-\nu_0^{\alpha} |t|},
  \label{dissipation1}
\end{equation}
where $\nu_{0}^{\alpha} \equiv \gamma_{\alpha}$, ${\bar c}_0^{\alpha}=\eta_{\alpha} \gamma_{\alpha}^{2}/2$, $c_{0}^{\alpha} \equiv \eta_{\alpha} \gamma_{\alpha}^{2} \cot( \beta \hbar \gamma_{\alpha}/2 )/2$
and $\nu_k^{\alpha} \equiv {2 \pi k}/{\beta \hbar}$,
\begin{equation}
  c_{k}^{\alpha} \equiv -{\eta_{\alpha} } \frac{4 \pi k  \gamma_{\alpha}^{2}}{(\beta \hbar  \gamma_{\alpha})^2-(2 \pi  k)^2}
\end{equation}
for $k>0$.  Under the high-temperature condition of $\beta \hbar \gamma_{\alpha} \ll 2$, the symmetrized correlation function is expressed as
$\langle \{\hat{\Omega}_{\alpha}(t), \hat{\Omega}_{\alpha}(0) \} \rangle_{\mathrm{B}}/2 = {\eta_{\alpha} \gamma_{\alpha}}\mathrm{e}^{-\gamma_{\alpha} |t|}/{\beta \hbar}$: The noise correlation function in the stochastic theory agrees with the high temperature limit of this fluctuation term. Thus, the results from the stochastic theory that includes the Kubo-Toyabe theory can be obtained from the HEOM approach for $J_{\alpha}(\omega) = \Delta_{\alpha}^2 \beta \hbar {\gamma_{\alpha} \omega}/({\omega^2 + \gamma_{\alpha}^{2}})$ with the dissipation term ignored.\cite{tanimura1989}

\subsection{HEOM approach}
The HEOM are the equations of motion that allow us to simulate the irreversible dynamics of the system through the fluctuation and dissipation given by Eqs. \eqref{fluctuation1} and \eqref{dissipation1}  in non-perturbative and non-Markovian manners at finite temperature.\cite{tanimura1989,TanimuraPRA90,TanimuraPRA91,ishizaki2005, ishizaki2006,tanimura2006,Tanimura2014,Tanimura2015} In this formalism, the effects of higher--order non-Markovian system-bath interactions are mapped into the hierarchical elements of the reduced density matrix. This formalism is valuable because it can be used to treat not only strong system--bath coupling but also quantum coherence (quantum entanglement) between the system and the bath, which is essential for studying a system subject to a time-dependent external force and nonlinear response functions.\cite{tanimura2006} Various analytical and numerical techniques have been developed for the HEOM approaches that allow us to study a complex system under quantum mechanically extreme conditions.
With the above described features, the HEOM, which were developed to bridge between the Markovian and perturbative quantum master equation theory and non-Markovian and non-perturbative but phenomenological SLE theory, exhibit wide applicability. The HEOM approach is ideal for extending the applicability of the Kubo-Toyabe low-field theory to various problems and physical conditions in a rigorous manner, and has been applied to spin relaxation problems.\cite{tanimura2006,Joutsuka2008,Tsuchimoto2015,Nakamura2018} Here, we investigate the $\mu$SR problem using the HEOM approach.

In the case of the three-dimensional spin-Boson model, the HEOM is given by\cite{ishizaki2005, ishizaki2006,tanimura2006,Tanimura2014}

\begin{align}
  \frac{\partial}{\partial t} \hat{\rho}_{\bm{n}}(t) =
  & - i \hat{\mathcal{L}} \hat{\rho}_{\bm{n}}(t) \; - \sum_{\alpha = x, y, z} \left[ \sum_{k = 0}^{K_{\alpha}} n_k^{\alpha} \nu_k^{\alpha} + \hat{\Xi}^{\alpha} \right] \hat{\rho}_{\bm{n}}(t) \notag \\[+4pt]
  & - \sum_{\alpha = x, y, z} \sum_{k = 0}^{K_{\alpha}} \hat{\Phi}^{\alpha}\hat{\rho}_{\bm{n} + \bm{e}_{k}^{\alpha}}(t)
  - \sum_{\alpha = x, y, z} \sum_{k = 0}^{K_{\alpha}} n_k^{\alpha} \hat{\Theta}_k^{\alpha} \hat{\rho}_{ \bm{n} - \bm{e}_k^{\alpha} }(t),
\label{HEOM}
\end{align}
where $i\hat{\mathcal{L}} \equiv {i}\hat{H}_{\mathrm{S}}^{\times}/\hbar$ and we introduce the set of hierarchy elements $\bm{n} \equiv \{ \bm{n}^x;\, \bm{n}^y; \,\bm{n}^z \}$ with
$\bm{n}^{\alpha}\equiv\{ n_0^{\alpha}, \cdots, n_{K_{\alpha}}^{\alpha}\}$ for $\alpha = x, y, $ and $z$, and the unit vector along the $k$th element in the $\alpha$ direction expressed as $\pm\bm{e}_k^{\alpha}$ that changes the index of the $n_k^{\alpha}$ element as $n_k^{\alpha}\pm1$.
Here, $n_0^{\alpha}$ is the element for $\gamma^{\alpha}$, whereas $n_k^{\alpha}$ for $k \ge 1$ are the elements for the Matsubara frequencies $\nu_k^{\alpha}$, respectively,  in the $\alpha$ direction.
The $\alpha$th bath-induced relaxation operators are defined as $\hat{\Phi}^{\alpha} \equiv i \hat{V}_{\alpha}^{\times}$,  $\hat{\Theta}_k^{\alpha} \equiv i c_k^{\alpha} \hat{V}_{\alpha}^{\times}$,
\begin{align}
  \hat{\Theta}_0^{\alpha} \equiv -\bar{c}_0^{\alpha} \hat{V}_{\alpha}^{\circ} + i c_0^{\alpha} \hat{V}_{\alpha}^{\times},
  \label{Theta0}
\end{align}
and
\begin{align}
  \hat{\Xi}^{\alpha} \equiv \left[ -\sum_{k=1}^{K_{\alpha}} \frac{c_k^{\alpha}}{\nu_{k}^{\alpha}} +  \left( \frac{\eta_{\alpha}}{\beta \hbar} - \frac{c_0^{\alpha}}{\gamma_{\alpha}} \right)  \right] \hat{V}_{\alpha}^{\times} \hat{V}_{\alpha}^{\times},
  \label{Xi}
\end{align}
where we have introduced the hyperoperator notation
$\hat{\mathcal{O}}^{\times} \hat{f} \equiv  [ \hat{\mathcal{O}}, \hat{f} ]$ and $\hat{\mathcal{O}}^{\circ} \hat{f} \equiv \{ \hat{\mathcal{O}}, \; \hat{f} \}$
for any operator $\hat{\mathcal{O}}$ and operand operator $\hat{f}$.  The hierarchy of equations of motion introduced above continues to infinity, which is not easy to solve numerically.  To truncate Eq.(\ref{HEOM}), we introduce the terminator\cite{TanimuraPRA91,ishizaki2005, tanimura2006}
\begin{equation}
  \frac{\partial}{\partial t} \hat{\rho}_{\bm{n}}(t)
  \simeq - i \hat{\mathcal{L}} \hat{\rho}_{\bm{n}}(t) - \sum_{\alpha = x, y, z} \hat{\Xi}^{\alpha} \hat{\rho}_{\bm{n}}(t),
\end{equation}
which is valid for the integers $n_0^{\alpha}, \cdots, n_{K_{\alpha}}^{\alpha}$satisfying
\begin{equation}
  \sum_{k=0}^{K_{\alpha}} n_{k}^{\alpha} \gg \frac{\omega_c}{\min(\gamma_{\alpha}, \nu_{1}^{\alpha})}.
\end{equation} In the high temperature case, the HEOM reduces to\cite{tanimura1989, tanimura2006}
\begin{align}
 \frac{\partial}{\partial t}
 \hat{\rho}_{\bm{n}}(t) =& - \left(i \hat{\mathcal{L}} \; + \sum_{\alpha = x, y, z} n_0^{\alpha} \gamma_{\alpha} \right) \hat{\rho}_{\bm{n}}(t)
- \sum_{\alpha = x, y, z} \hat{\Phi}^{\alpha} \hat{\rho}_{\bm{n} + \bm{e}^{\alpha}}(t) \nonumber \\
&- \sum_{\alpha = x, y, z} n_{0}^{\alpha}\hat{\Theta}_{0}^{\alpha} \hat{\rho}_{ \bm{n} - \bm{e}^{\alpha} }(t),
\label{HEOMhigh}
\end{align}
where $c_0^{\alpha}$ in Eq.\eqref{Theta0} is now approximated as $ c_0^{\alpha}=\eta_{\alpha} \gamma_{\alpha}/\beta\hbar$ and $\bm{n}$ reduces to
$\bm{n}=\{ n_0^x;  n_0^y;  n_0^z \}$ with  $\bm{e}^{\alpha}\equiv\bm{e}_0^{\alpha}$.
Through numerical integration of the equations, we can calculate $\mu$SR spectrum under any physical condition even under a time-dependent external force. In the Markovian limit $\gamma_{\alpha} \gg \omega_c$, the above equation further reduces to the master equation
\begin{align}
 \frac{\partial}{\partial t}
 \hat{\rho}(t) = - i \hat{\mathcal{L}}  \hat{\rho}(t)
- \sum_{\alpha = x, y, z}  \eta_{\alpha}  \hat{V}_{\alpha}^{\times}
\left(\frac{1}{\beta\hbar}\hat{V}_{\alpha}^{\times} - i \frac{\gamma_{\alpha}}{2} \hat{V}_{\alpha}^{\circ} \right) \hat{\rho}(t).
 \label{HEOMarkov}
\end{align}
In the case of regular NMR described by finite $\omega_0$, the HEOM can describe the $T_1$ and $T_2$ relaxation processes from the $x$ and $y$ baths, respectively, and the $T_2^{\dag}$ relaxation process from the z bath in the fast modulation limit without the rotating wave approximation (RWA). While the above equation is valid only in the high temperature case, the HEOM presented in Eq. \eqref{HEOM} is valid at any temperature under non-Markovian conditions.

The SLE can also obtained from Eq. \eqref{HEOMhigh} by assuming an extremely high temperature case by ignoring the term $ i {\gamma_{\alpha}} \hat{V}_{\alpha}^{\circ}/2$ and by rescaling $\Delta_{\alpha}^2 = \eta_{\alpha} \gamma_{\alpha} / \beta \hbar$ to obtain\cite{tanimura1989,tanimura2006}
\begin{align}
 \frac{\partial}{\partial t} \hat{\rho}_{\bm{n}}(t) =& - \left(i \hat{\mathcal{L}} \; + \sum_{\alpha = x, y, z} n_0^{\alpha} \gamma_{\alpha} \right) \hat{\rho}_{\bm{n}}(t)
- \sum_{\alpha = x, y, z} i \Delta_{\alpha}\hat{V}_{\alpha}^{\times} \hat{\rho}_{\bm{n} + \bm{e}^{\alpha}}(t) \nonumber \\
&- \sum_{\alpha = x, y, z} i n_0^{\alpha}  \Delta_{\alpha} \hat{V}_{\alpha}^{\times}
\hat{\rho}_{ \bm{n} - \bm{e}^{\alpha} }(t).
\label{SLE}
\end{align}
AAlthough the numerical cost of solving Eq. \eqref{SLE} is almost the same as that of solving Eq. \eqref{HEOMhigh}, we can solve the above equation in the same manner as Eq. \eqref{HEOMhigh} using the truncation scheme developed for the HEOM formalism.

\section{Results and Discussion}
\label{Results}
 We calculated the free induction decay signal of a spin polarization defined by $G_z(t)= \mathrm{Tr} \left\{\hat{\rho}(t) \cdot \hat{\sigma}_{z} \right\}$.  To reduce the computational costs, we constructed the HEOM using the Pad{\'e}-based expression for $c_{k}^{\alpha}$ and $\nu_{k}^{\alpha}$ instead of using the Matsubara--frequency--based expression.\cite{Hu2010,Hu2011,Ding2011} Numerical calculations were carried out to integrate Eq.~\eqref{HEOMhigh} using the fourth-order low-storage Runge-Kutta (LSRK4) method,\cite{LSRK42017,Ikeda2018CI} with a time step of $\delta t=0.01\times 10^{-2}$. We considered the factorized initial state with the 100\% polarized spin in the +z direction, i.e., $G_z(0)=1$, to account for the condition of the actual $\mu$SR measurement. By numerically integrating the SLE presented in Eq. \eqref{SLE}, we also calculated the stochastic results to illustrate the roles of the dissipation and the low temperature correction terms involved in the HEOM.
We first considered the case of the isotropic environment described by $\gamma_x=\gamma_y=\gamma_z=\gamma$ and $\eta_x=\eta_y=\eta_z=\eta$.

\subsection{Temperature effects: Interplay between fluctuation and dissipation}
\label{tempeffect}
\begin{figure}[t]
  \centering
  \includegraphics[scale = 0.5]{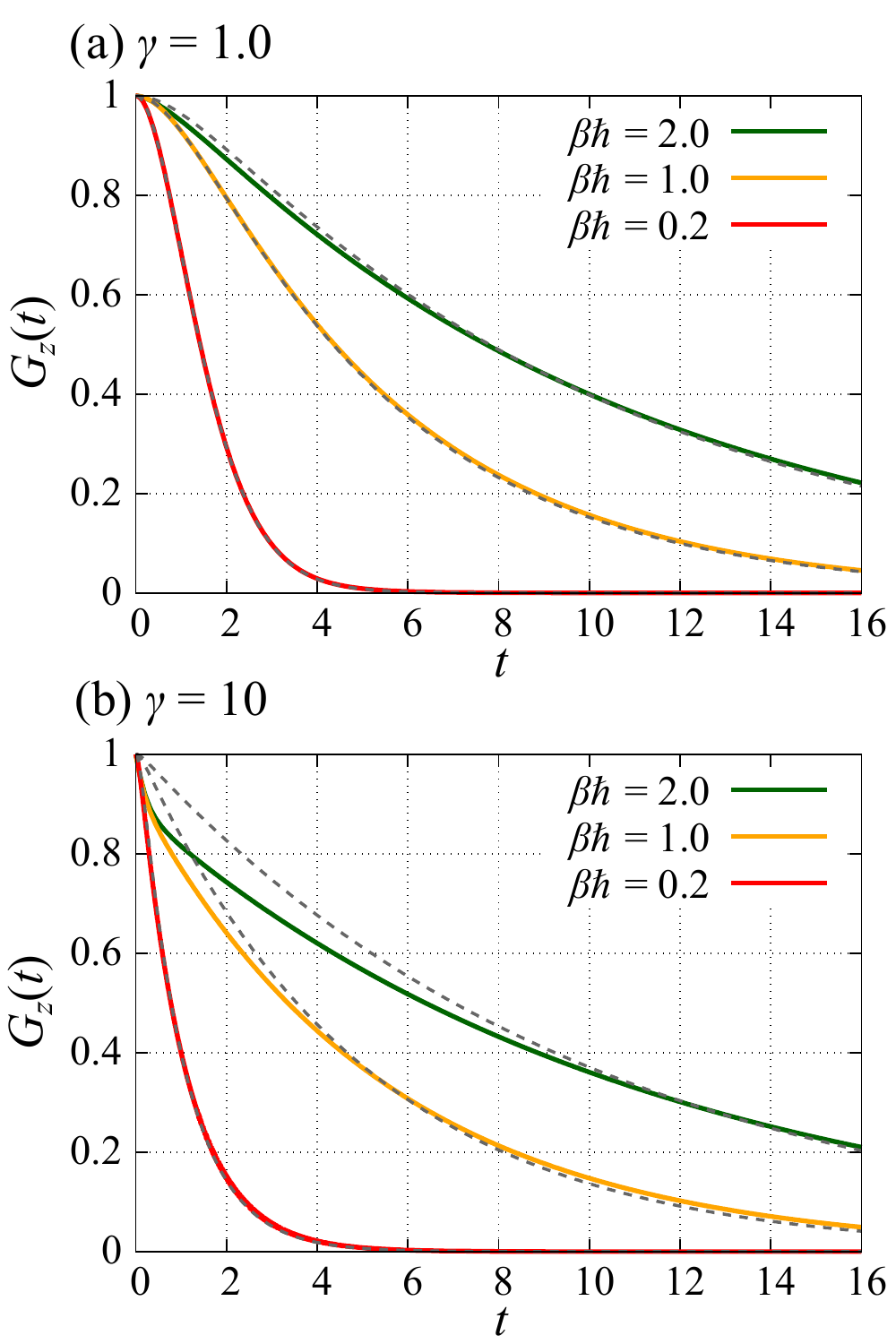}
  \caption{(Color online) $\mu$SR spectrum calculated using HEOM (solid curves) and SLE (dashed curves) under (a) intermediate modulation ($\gamma=1$) and (b) fast modulation ($\gamma=10$) conditions for weak coupling case ($\eta=0.1$) with various inverse temperatures $\beta\hbar=2.0$, 1.0, and 0.2. Because the SLE theory does not account for the temperature effects, we adjusted the amplitude of the stochastic modulation as $\Delta_{\alpha}^2 = \eta_{\alpha} \gamma_{\alpha} / \beta \hbar$.}
  \label{temperature}
\end{figure}
In Fig. \ref{temperature}, we show the temperature dependence of the $\mu$SR spectrum for the fixed coupling strength $\eta=0.1$ for two cases of the inverse noise correlation time (a) $\gamma=1$ and (b) $\gamma=10$. We compare the HEOM (solid curves) and SLE (dashed curves) results to study the role of the dissipation term by setting $\Delta_{\alpha}^2 = \eta_{\alpha} \gamma_{\alpha} / \beta \hbar$.
In the intermediate modulation case shown in Fig. \ref{temperature}(a), the HEOM and SLE results are all similar, whereas in the fast modulation case shown in Fig. \ref{temperature}(b), they are different in the low temperature cases. In the HEOM formalism, the high temperature condition is written as $\beta\hbar\gamma /2 < 1$. This implies that all the cases in Fig. \ref{temperature}(a) and the case $\beta\hbar=0.2$ in Fig. \ref{temperature}(b) are in the high temperature regime, where the HEOM reduce to Eq. \eqref{HEOMhigh}.
Because we adjusted the amplitude of the stochastic noise to fit the HEOM results, the difference in the HEOM results arises only from the dissipation term presented as the first term in Eq. \eqref{Theta0}, which becomes negligible for a small $\beta\hbar\gamma$ in comparison with the second term. This indicates that, by setting $\Delta_{\alpha}^2 = \eta_{\alpha} \gamma_{\alpha} / \beta \hbar$, we may explain the temperature dependence of the $\mu$SR spectrum within the framework of the stochastic theory under such conditions. When the temperature becomes very low, however, the signals calculated from the HEOM decay rapidly in comparison with those from the SLE. This difference is due to the time-irreversible dynamics of the spin described by the interplay of the fluctuation and dissipation, whereas the SLE includes dephasing only described by the fluctuation. Because the contribution of the dissipation term becomes large in the low temperature regime, the HEOM results decay more rapidly.

\subsection{Non-Markovian effects: Role of quantum thermal noise}
\label{nonMarkovi}
\begin{figure}[t]
  \centering
  \includegraphics[scale = 0.5]{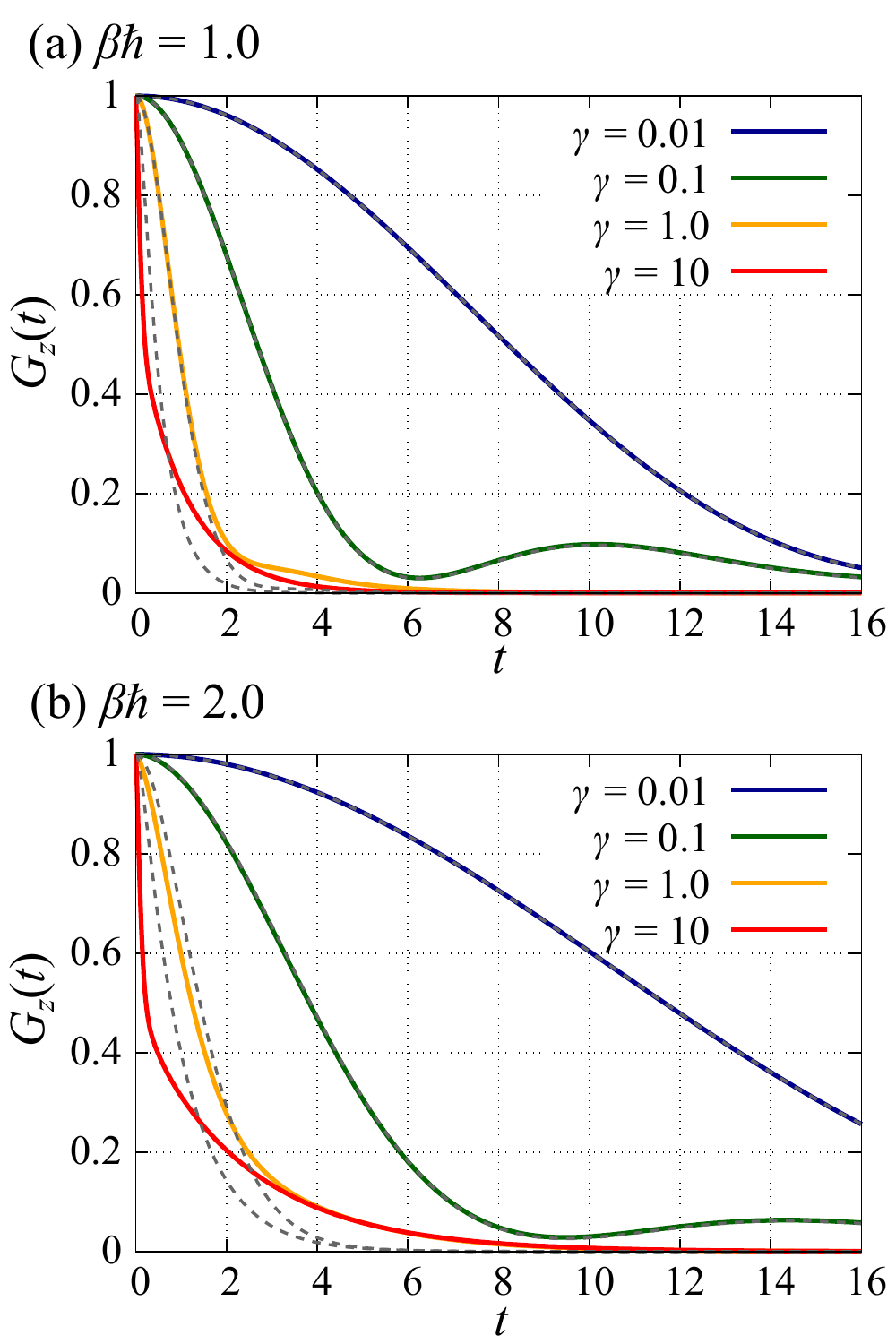}
  \caption{(Color online) $\mu$SR spectrum calculated using HEOM and SLE for weak coupling case ($\eta=0.1$) under (a) intermediate temperature ($\beta\hbar=1.0$) and (b) low temperature ($\beta\hbar=2.0$) conditions with various modulation rates $\gamma$=0.01, 0.1, 1, and 10. The solid and dashed curves represent the HEOM and corresponding SLE results, respectively.}
  \label{nonMarkov}
\end{figure}

Next, we investigate the effect of the noise correlation (non-Markovian effects) in the weak coupling case under the (a) intermediate temperature ($\beta\hbar=1.0$) and (b) low temperature ($\beta\hbar=2.0$) conditions.
Because the condition $\beta\hbar\gamma \le 0.5$ is maintained, the HEOM and SLE results exhibit similar Gaussian decay profiles for the slower modulation case as predicted from the stochastic theory.  Note that, when $\gamma$ is sufficiently small, both HEOM and SLE results exhibit a 1/3 tail that was predicted by the static limit of the Kubo-Toyabe theory.  The distinct feature of the HEOM results is observed in the fast modulation cases ($\gamma$=10) in Figs. \ref{nonMarkov}(a) and \ref{nonMarkov}(b): The signals calculated from the HEOM  decay more slowly than those calculated from the SLE after exhibiting a fast initial decay in the time period less than $1/\beta\hbar\approx1.0$ or 0.5.

While the fast decay is due to the population relaxation arising from the dissipation, the slow decay is due to the quantum dephasing arising from the quantum thermal noise. As illustrated in Eqs. \eqref{fluctuation1} and \eqref{dissipation1}, two types of non-Markovian noise are involved in the system dynamics: one is of mechanical origin characterized by the fluctuation ($c_{0}^{\alpha}\mathrm{e}^{-\nu_{0}^{\alpha} t}$)
and dissipation ($\bar{c}_{0}^{\alpha}\mathrm{e}^{-\nu_{0}^{\alpha} t}$) with $\nu_0^{\alpha}=\gamma$, and the other is of quantum thermal origin characterized by the fluctuation only ($c_{k}^{\alpha} \mathrm{e}^{-\nu_{k}^{\alpha}t}$ for $k\ge 1$) with $\nu_{k}^{\alpha} = 2\pi k/\beta\hbar$.
When $\gamma$ is much larger than $\nu_1$, the mechanical contribution with ${\rm e}^ {-\nu_0^{\alpha} t}$ vanishes after $t>1/\nu_1$, and the effects from the quantum thermal noise take place. The quantum thermal fluctuation exhibits a peculiar behavior in comparison with the mechanical fluctuation, because the amplitude of the noise becomes negative for a large a $\gamma$ [see Fig. 7(b) in Ref. ~\citen{tanimura2006}]. Thus, the signal obtained from the HEOM decays more slowely than that obtained from the SLE. Although the SLE is also a non-Markovian theory, this quantum thermal dephasing process can be described only from the numerically ``exact'' HEOM approach.

\subsection{Non-perturbative system-bath interactions}
\begin{figure}[t]
  \centering
  \includegraphics[scale = 0.5]{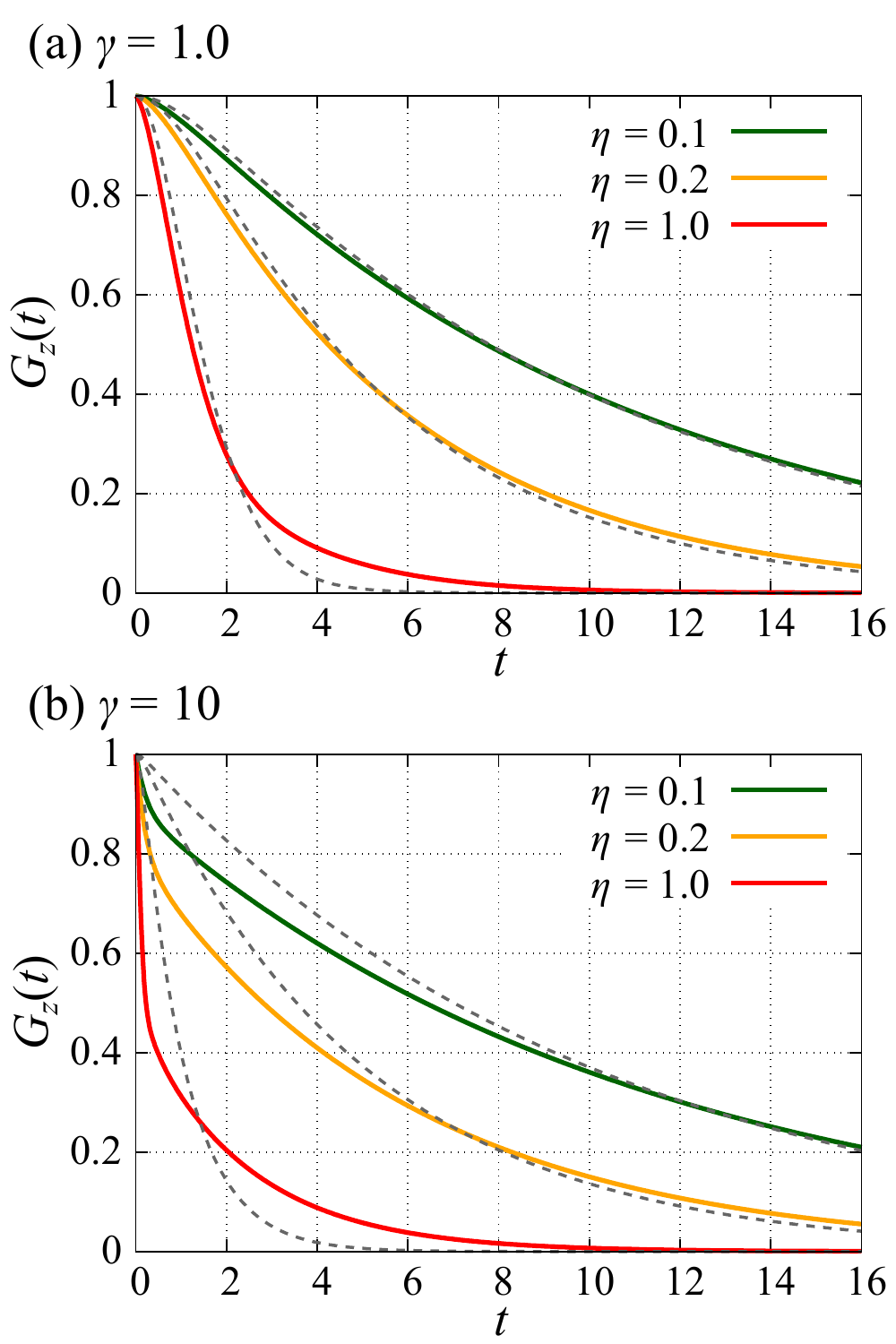}
  \caption{(Color online) $\mu$SR spectrum calculated for low temperature case ($\beta\hbar=2$) under (a) intermediate modulation ($\gamma=1.0$) and (b) fast modulation ($\gamma=10$) conditions with various coupling strengths $\eta$=0.1, 0.2, and 1. The solid and dashed curves represent the HEOM and stochastic results, respectively.}
\label{nonperturbative}
\end{figure}
We study the non-perturbative effects of the system--bath coupling by changing $\eta$. As illustrated in Fig. \ref{nonperturbative}(a), the differences between the HEOM and SLE results increase with coupling strength even in the intermediate modulation case. This is because, while the amplitudes of the fluctuation and dissipation are both proportional to the coupling strength, the relaxation arising from the dissipation plays a greater role than the dephasing arising from the fluctuation, owing to the time-irreversible nature of the relaxation. As depicted in Fig. \ref{nonperturbative}(b), such differences become prominent in the faster modulation case, as in the cases described in Sects. \ref{tempeffect} and \ref{nonMarkovi}. The time period of the initial decay decreases with increasing coupling strength, because the noise with $c_k{\mathrm{e}}^{-\nu_k^{\alpha} t}$ for a larger $k$ can interact with the system several times in this non-perturbative regime.

\subsection{Anisotropic effects of environment}

The HEOM formalism is ideal for studying a spin system under realistic conditions, because it allows the treatment of various anisotropic environments with any profile of noise correlation functions characterized by
\begin{equation}
  J_{\alpha, \alpha'}(\omega) = \sum_{j} g_j^{\alpha} g_j^{\alpha'}\; \delta(\omega - \omega_j^{\alpha\alpha'})
\end{equation}
for any combination of $\alpha, \alpha' =x,y, z$. This is because the HEOM formalism is based on the equations of motion approach. Below, we investigate the anisotropy effects of noise amplitudes and noise correlations. For this purpose, we consider the extremely high temperature case ($\beta\hbar=2.5\times 10^{-3}$) with the weak system-bath coupling $\eta=1.0\times 10^{-3}$. Thus, the HEOM and SLE results become almost identical for $\Delta_{\alpha}^2 = \eta_{\alpha} \gamma_{\alpha} / \beta \hbar$.

\begin{figure}[h]
  \centering
  \includegraphics[scale = 0.42]{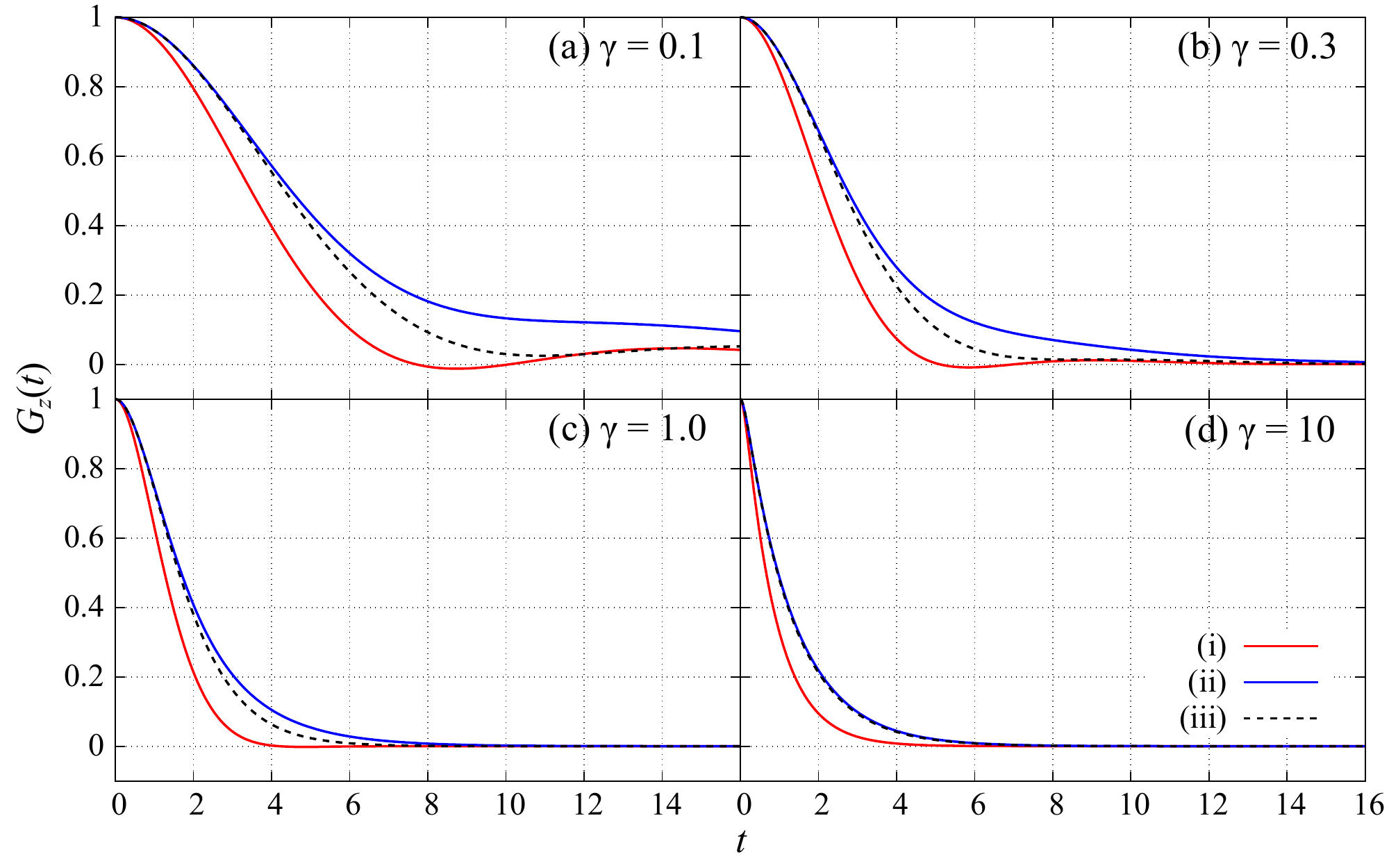}
  \caption{(Color online) $\mu$SR spectrum calculated for (i) $x$ anisotropic (red curves), (ii) $z$ anisotropic (blue curves), and (iii) isotropic (dashed curves) cases for (a) $\gamma=$ 0.1, (b) 0.3, (c) 1.0, and (d) 10. }
\label{anisotropic}
\end{figure}

\subsubsection{Anisotropic noise amplitudes}

We first study the effects of the anisotropic system-bath coupling strength expressed as
\begin{equation}
  \eta_{x} = a \eta, \;\; \eta_{y} = b \eta, \;\; \eta_{z} = c \eta,
\end{equation}
where $a$, $b$, and $c$ are the anisotropic constants. While the analysis of anisotropic effects was limited in the static case on the basis of the Kubo-Toyabe theory,\cite{aoyama1992, aoyama1993} there is no technical limitation from the HEOM approach for such problems, because we are only integrating the HEOM.

In Fig. \ref{anisotropic} we present the signals for the (i) $x$ anisotropic ($a=1$, $b=2$, and $c=1$), (ii) $z$ anisotropic ($a=1$, $b=1$, and $c=2$), and (iii) isotropic ($a=1$, $b=1$, and $c=1$) cases for various $\gamma$ values.
The other parameters are fixed as $\beta\hbar=2.5\times 10^{-3}$ and $\eta=1.0\times 10^{-3}$. In Figs. \ref{anisotropic}(a) -- \ref{anisotropic}(c), the signal decays more rapidly in the $x$ anisotropic case than in the isotropic case, whereas the signal decays more slowly in the $z$ anisotropic case than in the isotropic case.
This is because the $\hat \sigma_x$ operator causes the longitudinal ($T_1$) relaxation, whereas the $\hat \sigma_z$ operator causes not a relaxation but a dephasing ($T_2^{\dag}$) for the $z$-polarized spin. When $\gamma$ increases, the spin distribution approaches the equilibrium value owing to the relaxation. In the fast modulation case shown in Fig. \ref{anisotropic}(d), the signal decays more rapidly than in the case shown in Fig. \ref{anisotropic}(c), because the effective coupling strength becomes larger for a large $\gamma$ owing to the factor $\gamma_{\alpha}^2/(\gamma_{\alpha}^2 + \omega_c^2)$, where $\omega_c$ is the characteristic frequency of the system dynamics. The $z$ anisotropic results become similar to the isotropic case, because when the spin element in the $z$ direction becomes small, the effects of dephasing in the $z$ direction also become minimal.

\subsubsection{Anisotropic noise correlation}
\begin{figure}[t]
  \centering
  \includegraphics[scale = 0.42]{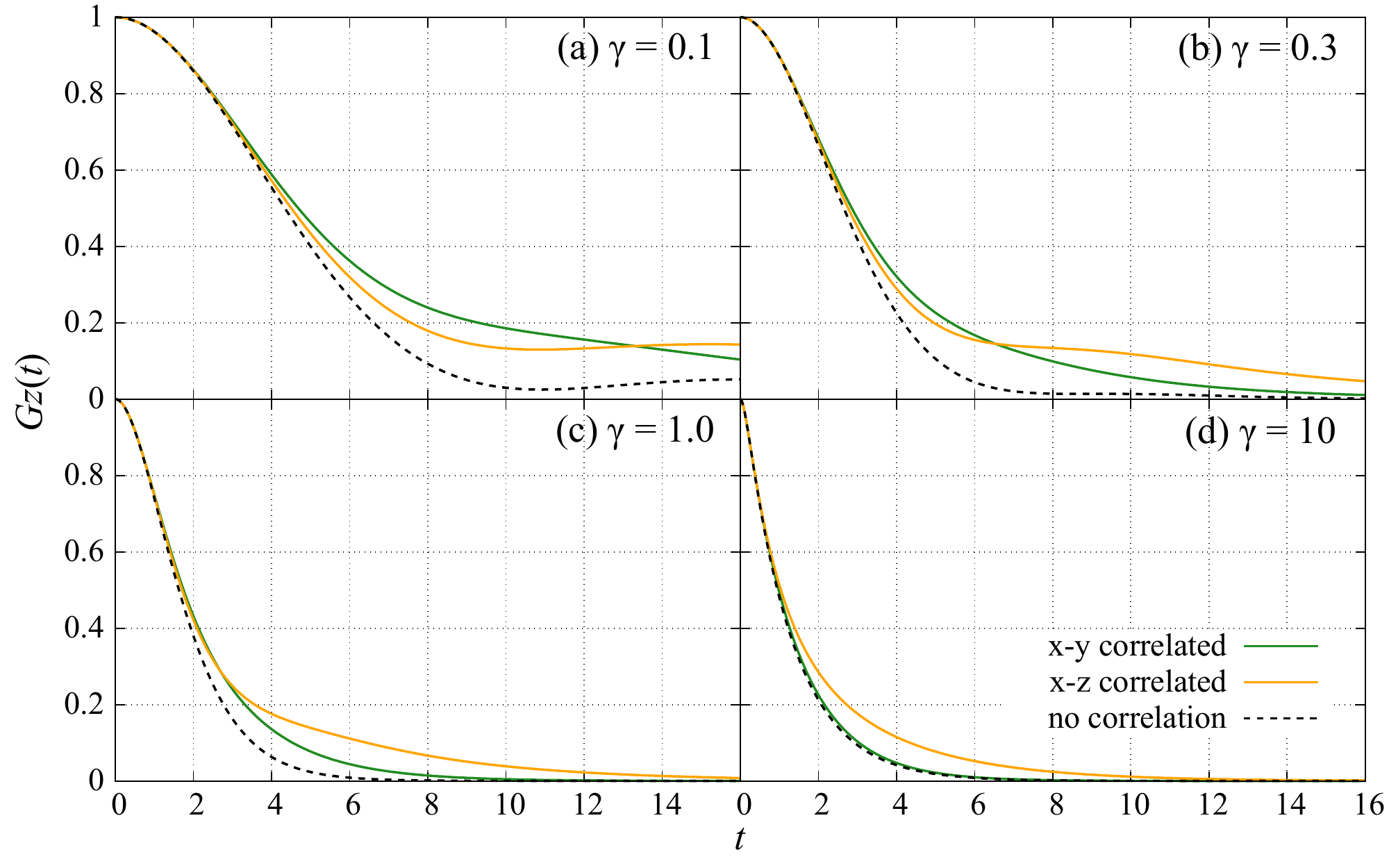}
  \caption{(Color online) $\mu$SR spectrum calculated for (i) $x-y$ correlated (green curves), (ii) $x-z$ correlated (orange curves), and (iii) isotropic (dashed curves) cases for (a) $\gamma=$ 0.1, (b) 0.3, (c) 1.0, and (d) 10.}
  \label{f6}
\end{figure}

Finally, we consider the case that some of the three-dimensional baths are correlated. Such a model was developed to analyze the noise correlation of different vibrational modes by two-dimensional infrared spectroscopy.\cite{Tokmakoff2003,ishizaki2007}
Here, we consider the (i) $x-y$ correlated [$\hat{V}_{x-y} = (\hat{\sigma}_{x} + \hat{\sigma}_{y})/2$ and $\hat{V}_{z} =  \hat{\sigma}_{z}$],
and (ii) $x-z$ correlated [$\hat{V}_{x-z} = (\hat{\sigma}_{x} + \hat{\sigma}_{z})/2$ and $\hat{V}_{y} = \hat{\sigma}_{y}$] cases for the bath Hamiltonian, Eq. \eqref{SBINT}, with (i) $\alpha= x-y$, and $z$, and (ii) $\alpha= x-z$, and $y$. These results are compared with (iii) the isotropic case $\alpha= x$, $y$, and $z$.

In Fig. \ref{f6}, $\mu$SR spectra are presented for various inverse noise correlation times $\gamma$ = (a) $0.1$, (b) $0.3$, (c) $1$, and (d) $10$. The characteristic feature of the present results is illustrated from the fluctuation term in Eq. \eqref{HEOMarkov} expressed as $(\eta_{\alpha}/{\beta\hbar})  \hat{V}_{\alpha}^{\times}\hat{V}_{\alpha}^{\times}$.
In the $x-y$ correlation case, this term is expressed as ${\hat V}_{x-y}^{\times} {\hat V}_{x-y}^{\times} = (\hat{\sigma}_{x}^{\times 2} + {\hat C}_{x-y} + \hat{\sigma}_{y}^{\times 2})/4$,
where ${\hat C}_{x-y} = (\hat{\sigma}_{x}^{\times} \hat{\sigma}_{y}^{\times} + \hat{\sigma}_{y}^{\times} \hat{\sigma}_{x}^{\times})$.
In the slow modulation case shown in Fig. \ref{f6}(a), the movements of the spin in the $x$ and the $y$ directions are not correlated, and the contribution from $\hat{C}_{x-y}$ becomes small. Thus, the longitudinal ($T_1$) relaxation becomes weaker than that in the isotropic case.
In the fast modulation case shown in Fig. \ref{f6}(d), however, the contribution from ${\hat C}_{x-y}$ becomes similar to that of $\hat{\sigma}_{x}^{\times 2}$ and $\hat{\sigma}_{y}^{\times 2}$,
and thus we have $\hat{V}_{x-y}^{\times} \hat{V}_{x-y}^{\times} \approx {\hat V}_{x}^{\times}{\hat V}_{x}^{\times} + {\hat V}_{y}^{\times}{\hat V}_{y}^{\times}$, which leads to the $x-y$ result becoming similar to the isotropic one.
In the $x-z$ correlation case, we have ${\hat V}_{x-z}^{\times} {\hat V}_{x-z}^{\times} < ({\hat V}_{x}^{\times} {\hat V}_{x}^{\times} + {\hat V}_{z}^{\times} {\hat V}_{z}^{\times})$ under the slow modulation condition, whereas the contribution from the $y$ direction does not change. Thus the $x-z$ signal shown in Fig. \ref{f6}(a) decays more slowely than that in the isotropic case, whereas it still decays more rapidly than the $x-y$ correlated signal. Under the fast modulation condition shown in Fig. \ref{f6}(d), the decay of the signal in the $x-z$ correlated case is slow, because, for the $z$-polarized spin, the ${\hat C}_{x-z}$ contribution remains small even under the fast modulation condition.

The above results indicate that the $\mu$SR spectrum is sensitive to the anisotropic effects of the environment, which should be detected experimentally in accordance with the theoretical analysis.

\section{Conclusions}
\label{conclusion}
As illustrated in this paper, the HEOM approach has distinct features for the analysis of $\mu$SR and near-zero-field NMR spectra. First, while the stochastic approach can treat the high-temperature Markovian case only, the HEOM approach can treat the realistic non-Markovian noise arising from complex environments, such as nanomaterials, proteins, and a spin lattice in different magnetic ordered phases. This is because the HEOM are constructed on the basis of a fairly complex system-bath Hamiltonian: It is also possible to construct a simulation model on the basis of a molecular dynamics simulation.\cite{Renger06, KramerAspu13, KramerAspu14, Coker2016, Ueno2020}
Second, because the HEOM is a dynamical theory, we can easily and clearly identify the roles of the system-bath interaction, noise correlation time, and heat-bath temperature. Third, because the HEOM approach is the equations of motion approach, there is no difficulty in taking into account the effects of a time-dependent external field. This allows us to calculate multi-dimensional near zero-field NMR signals for various pulse sequences. In addition, we can include the effects of nonthermal local environmental modes explicitly as the time-dependent external field, whereas the other thermal effects are taken into account using the hierarchical structure. For numerical integration, we can employ a complex quantum system, such as a spin chain\cite{Joutsuka2008} or a spin lattice\cite{Tsuchimoto2015,Nakamura2018} as the main system instead of a simple spin system.

In conclusion, the present formalism provides a powerful means of analyzing $\mu$SR and various NMR measurements for the study of environmental effects. It is also possible to use the HEOM theory to investigate other scattering and spectroscopic measurements, which include neutron scattering, electron paramagnetic resonance (EPR), and M{\" o}ssbauer measurements.\cite{tanimura2006} All of the possibilities mentioned above can be carried out as future studies upon request.

\section*{Acknowledgments}
The financial support from the Kyoto University Foundation is acknowledged.

\appendix

\section{HEOM for Brownian Spectral Distribution}
By extending the hierarchy, we can derive HEOM for the Brownian spectral distribution given by\cite{TanimuraMukamelJPSJ94,TanakaJPSJ09}
\begin{equation}
  J_{\alpha}(\omega) = \frac{\eta_{\alpha}}{\pi} \frac{\gamma_{\alpha}^2 \omega_{0\alpha}^{2} \omega }{(\omega_{0\alpha}^{2}-\omega^{2})^2 + \gamma_{\alpha}^2 \omega^2 },
  \label{Disposc}
\end{equation}
where $\omega_{0\alpha}$ is the frequency of the local mode, $\gamma_{\alpha}$ is the inverse correlation time of the noise,  and $\eta_{\alpha}$ is the coupling strength of the environment. Because the spectral density has two poles in the upper half-plane, $i\nu_{0}^{\alpha} \equiv i(\gamma_{\alpha}/2 - i\zeta_{\alpha})$ and $i\bar{\nu}_{0}^{\alpha} \equiv i(\gamma_{\alpha}/2 + i\zeta_{\alpha})$,
where $\zeta_{\alpha} \equiv \sqrt{\omega_{0\alpha}^{2} -\gamma_{\alpha}^{2}/4}$,  both the symmetrized and anti-symmetrized correlation functions are expressed as linear functions of
$\mathrm{e}^{-(\gamma_{\alpha}/2 - i \zeta_{\alpha})t}$ and $\mathrm{e}^{-(\gamma_{\alpha}/2 + i \zeta_{\alpha})t}$, in addition to Matsubara frequency terms, as defined earlier. Here and for all previously defined abbreviations throughout. Thus, we can construct the HEOM by evaluating the time derivative of the reduced density matrices as \cite{TanakaJPSJ09, TanakaJCP10}
\begin{align}
  \frac{\partial}{\partial t} \hat{\rho}_{\bm{n}}(t)
  =& - i \hat{\mathcal{L}} \hat{\rho}_{\bm{n}}(t)- \sum_{\alpha = x, y, z} \left[ n_{0}^{\alpha} \nu_{0}^{\alpha} + \bar{n}_{0}^{\alpha} \bar{\nu}_{0}^{\alpha}
  +\sum_{k=1}^{K_{\alpha}} n_{k}^{\alpha} \nu_{k}^{\alpha} - \hat{\Xi}^{\alpha}
  \right]  \hat{\rho}_{\bm{n}}(t)  \notag \\
  &- \sum_{\alpha=x,y,z} \left[ \hat{\Phi}^{\alpha}\hat{\rho}_{\bm{n}+\bm{e}^{\alpha}}(t)
   + n_{0}^{\alpha} \hat{\Theta}_{-}^{\alpha} \hat{\rho}_{\bm{n}-\bm{e}^{\alpha}} (t) \right] \notag \\
  &- \sum_{\alpha=x,y,z} \left[ \hat{\Phi}^{\alpha}\hat{\rho}_{\bm{n}+\bar{\bm{e}}^{\alpha}}(t)
   + \bar{n}_{0}^{\alpha} \hat{\Theta}_{+}^{\alpha} \hat{\rho}_{\bm{n}-\bar{\bm{e}}^{\alpha}} (t) \right] \notag \\
  &- \sum_{\alpha=x,y,z} \sum_{k=1}^{K_{\alpha}} \hat{\Phi}^{\alpha} \hat{\rho}_{\bm{n}+\bm{e}_{k}^{\alpha}}(t)  \notag \\
  &- \sum_{\alpha=x,y,z} \sum_{k=1}^{K_{\alpha}} n_{k}^{\alpha} \hat{\Psi}_{k}^{\alpha} \hat{\rho}_{\bm{n}-\bm{e}_{k}^{\alpha}}(t),
 \label{EOM}
\end{align}
where $\hat{\Phi}^{\alpha}=i\hat{V}_{\alpha}^{\times}$, $\hat{\Theta}_{\pm}^{\alpha} = - {\bar c}_{\pm}^{\alpha}\hat{V}_{\alpha}^{\circ} + i c_{\pm}^{\alpha} \hat{V}_{\alpha}^{\times}$,
$ \hat{\Psi}_{k}^{\alpha} = i c_{k}^{\alpha} \hat{V}_{\alpha}^{\times}$, and $\hat{\Xi}^{\alpha} = \hat{\Phi}^{\alpha} \sum_{k=K_{\alpha}+1}^{\infty} \hat{\Psi}_{k}^{\alpha}$
with  ${\bar c}_{\pm}^{\alpha}=\mp i{\eta_{\alpha}\omega_{0\alpha}^{2}}/{4\zeta_{\alpha}}$,
\begin{equation}
  c_{\pm}^{\alpha} = \mp \frac{\eta_{\alpha}\omega_{0\alpha}^{2}}{4\zeta_{\alpha}} \coth \left\{ \frac{\beta\hbar}{2} \left(i\frac{\gamma_{\alpha}}{2} \mp \zeta_{\alpha} \right) \right\},
\end{equation}
and
\begin{equation}
  c_{k}^{\alpha} = -\frac{2\eta_{\alpha}\omega_{0\alpha}^{2}}{\beta\hbar} \frac{\gamma_{\alpha} \nu_{k}^{\alpha}}{(\omega_{0\alpha}^{2}+ {\nu_{k}^{\alpha}}^{2})^{2}-\gamma_{\alpha}^{2} {\nu_{k}^{\alpha}}^2}.
  \label{Psi}
\end{equation}
The hierarchical elements $\bm{n}\equiv\{ \bm{n}^x;\,  \bm{n}^y; \bm{n}^z \}$ are now defined by
$\bm{n}^{\alpha} \equiv \{ {n}_0^{\alpha},  {\bar n}_0^{\alpha}, {n}_1^{\alpha}, \cdots, n_{K_{\alpha}}^{\alpha}\}$,
where $n_{0}^{\alpha}$ and $\bar{n}_{0}^{\alpha}$ are the elements for $\nu_{0}^{\alpha}$ and $\bar{\nu}_{0}^{\alpha}$ for $\alpha = x, y, $ and $z$, respectively. The unit vectors in the $\alpha$ direction, which change the indexes of the $n_0^{\alpha}$ and ${\bar n}_0^{\alpha}$ elements as $n_0^{\alpha}\pm1$ and ${\bar n}_0^{\alpha}\pm1$, are expressed as $\pm\bm{e}^{\alpha}$ and $\pm {\bm{\bar e}}^{\alpha}$, respectively, whereas the other unit vectors $\bm{e}_{k}^{\alpha}$ are defined in the same manner as Eq. \eqref{HEOM}.

For the condition $n_{0}^{\alpha} + \bar{n}_{0}^{\alpha} + \sum_{k=1}^{K_{\alpha}} n_{k}^{\alpha} \gg \omega_c / \min(\gamma_{\alpha}/2, \nu_{1}^{\alpha})$,
this infinite hierarchy can be truncated by the terminator as
\begin{equation}
  \frac{\partial}{\partial t} \hat{\rho}_{\bm{n}}(t)
  \simeq - i \hat{\mathcal{L}}\hat{\rho}_{\bm{n}}(t) - \sum_{\alpha=x,y,z} \left[ i(-n_{0}^{\alpha}+\bar{n}_{0}^{\alpha}) \zeta_{\alpha} - \hat{\Xi}^{\alpha} \right]\hat{\rho}_{\bm{n}}(t).
  \label{terminator3}
\end{equation}
In the high temperature case, the above equations reduce to\cite{TanimuraMukamelJPSJ94}
\begin{align}
  \frac{\partial}{\partial t} \hat{\rho}_{\bm{n}}(t)
  =& - i \hat{\mathcal{L}} \hat{\rho}_{\bm{n}}(t)
  - \sum_{\alpha=x,y,z} \left( n_{0}^{\alpha} \nu_{0}^{\alpha} + \bar{n}_{0}^{\alpha} \bar{\nu}_{0}^{\alpha} \right) \hat{\rho}_{\bm{n}}(t) \notag \\
  &- \sum_{\alpha=x,y,z} \left[ \hat{\Phi}^{\alpha} \hat \rho_{\bm{n}+\bm{e}^{\alpha}} (t) + n_{0}^{\alpha} \hat{\Theta}_{-}^{\alpha} \hat{\rho}_{\bm{n}-\bm{e}^{\alpha}}(t) \right] \notag \\
  &- \sum_{\alpha=x,y,z} \left[ \hat{\Phi}^{\alpha} \hat \rho_{\bm{n}+\bar{\bm{e}}^{\alpha}} (t)
  - \bar{n}_{0}^{\alpha} \hat{\Theta}_{+}^{\alpha} \hat{\rho}_{\bm{n}-\bar{\bm{e}}^{\alpha}}(t) \right],
  \label{eq:dispmast}
\end{align}
where
\begin{equation}
  \hat \Theta_{\pm}^{\alpha} =
  \pm \frac{\eta_{\alpha}}{4\zeta_{\alpha}} \left( \frac{-\gamma_{\alpha} + 2i\zeta_{\alpha}}{\beta\hbar} \hat{V}_{\alpha}^{\times}
  + i \omega_{0\alpha}^{2} \hat{V}_{\alpha}^{\circ} \right).
  \label{eq:thetapm}
\end{equation}

\end{document}